\documentclass[12pt]{article}
\usepackage{graphicx}

\begin{document}

\title{Statistical Signatures in  Times of Panic: Markets  as a Self-Organizing System}
\author{Lisa Borland  \\  Evnine  and Associates, Inc.\\456 Montgomery Street, 
Suite 800, San Francisco, CA 94104,  USA \\ lisa@evafunds.com}
\maketitle

\abstract{ 
We study  properties of the cross-sectional distribution of returns. A significant anti-correlation between dispersion and cross-sectional kurtosis is found such that dispersion is high but kurtosis is low in panic times, and the opposite in normal times. The co-movement of stock returns also increases in panic times. We define a simple statistic $s$, the normalized sum of signs of returns on a given day, to capture the degree of correlation in the system. $s$ can be seen as  the order parameter of the system because if $s= 0$ there  is no correlation (a disordered state), whereas for $s \ne 0$ there is  correlation among stocks (an ordered state).  We make an analogy to non-equilibrium phase transitions and  hypothesize that financial markets undergo self-organization when the external volatility perception rises above some critical value. Indeed, the distribution of $s$ is unimodal  in normal times, shifting to  bimodal  in times of panic. This is consistent with a  second order phase transition. Simulations of a joint stochastic process for stocks use a multi timescale process in the temporal direction and an equation for the order parameter $s$ for the dynamics of the cross-sectional correlation.  Numerical results  show good qualitative agreement with the stylized facts of real data, in both normal and panic times. }

\section{Introdcution}

One of the big challenges in modeling financial time series lies in the fact that we cannot run experiments; we simply have one single unique realization of history. We cannot control the environment at will. From this point of view, the recent financial crisis is a gift to quantitative analysts, in particular those of us who wish to understand more about the underlying dynamics of markets. In this paper we  explore statistical signatures of market panic, finding some interesting and until now - to our knowledge - undocumented properties. We then propose a joint stochastic process to describe the behavior of stocks across time as well as their interactions at each time point. We postulate a simple model of self organization, and claim that there is evidence of a spontaneous phase transition at the onset of panic.

The outline of this paper is a brief review of the known stylized facts observed in time-series of returns, as well as a  discussion of various plausible models of that temporal behavior, though we later shall focus on a multi-timescale feedback model which is seen to capture most known statistical properties. We then delve into some empirical observations of the cross-sectional behavior of markets, which is  followed by  a theoretical section which aims at presenting a joint stochastic process to describe the interaction of stocks across time. At that point we introduce the notion  of a particular macroscopic observable intimately related to the correlations in the system as an order parameter, borrowing concepts from physics to motivate what we believe to be evidence of self-organization from a disordered to ordered state  in market dynamics. Finally we run numerical simulations and find good qualitative agreement with empirical observations. A discussion of the results and a look at future research directions conclude our current work.

\section{Returns across time}

The stylized facts of financial instruments  across time is quite well known and especially in the past decade and a half  has been studied quite intensely. Perhaps the most striking feature is that returns calculated over time scales ranging from minutes to weeks are well fit by a power law distribution with the tail index 3 \cite{stanley, jpbouchaud}. Several classes of distributions can be fit to these  densities of returns but  for example a Student-t with about 4 degrees of freedom (equivalently a Tsallis distribution with q = 1.4)  is quite a good choice for daily returns \cite{tsallis, borland1}. A plot of such a fit for  the distribution of daily returns of the Dow Jones index is shown in Figure (1). As the time scale increase, the power law property persists, only slowly decaying to a Gaussian \cite{stanley}. Another interesting property is that volatility itself tends to cluster. There are periods of high volatility followed by quieter periods. This indicates that there is memory in the volatility process. In fact, an auto-correlation calculation shows that  this memory decays as $\tau^{-.3}$ where $\tau$ is the time scale \cite{jpbouchaud}. Furthermore, a multi-fractal analysis of the volatility shows that it is self-similar \cite{jpbouchaud}. This means that the volatility clustering occurs on all timescales, intraday and daily  for example. In addition, the distribution of volatility is well-modeled by a log-normal or inverse gamma distribution \cite{jpbouchaud}. On top of these features is the striking fact that financial time series are not time invariant; if you flip the order of the time series and calculate  future volatility conditioned on past realized volatility, you will find an asymmetry. This indicates that there is causality - the future volatility depends on the past \cite{zumbach,Mandelbrot}. Another asymmetric feature is the so-called leverage effect: large negative returns tend to precede higher volatility \cite{leverageeffect}.  These statistical features all  appear to be rather universal in the sense that they can be found for a variety of financial instruments (stocks and currencies for example) as well as in different geographic regions and at different periods in time.

While many models have been proposed to model the dynamics of returns, there are not many that can capture all of the stylized facts. The simplest model is that of Bachelier,
later made famous in a slight modification by Black and Scholes (BS) \cite{Black&Scholes,Merton}. That model assumes that log returns are driven by a Brownian motion with constant drift and constant volatility. Clearly this model is a mis-specification because the resulting  distribution of log returns is Gaussian, and there is no mechanism for volatility clustering or memory. Nevertheless that model has gained wide recognition because of its analytic tractability especially when used as a basis to price derivative instruments such as options. Modifications to the BS model include stochastic volatility models \cite{carr,heston}, for example the Heston model \cite{heston}, where  the volatility itself is assumed to follow a mean reverting stochastic process. Such models introduce an additional source of randomness, and do not  always capture the correct statistical properties of real returns, but lend some analytic tractability to certain problems. Another class of models which seem quite promising are constituted by what we shall refer to as  statistical feedback processes.  A few years ago we proposed  a non-Gaussian statistical feedback process  where the volatility depends on the probability of the returns themselves \cite{borland1,borland2,borland3}. That model captured many of the stylized facts including the persistent power-law distribution of returns and was used successfully to derive option pricing formulae. However the model is non-stationary in the sense that returns are always calculated with respect to a particular initial time and initial price; a generalization of that model to include feedback over multiple timescale was then developed and we found that this model captured all the known stylized facts \cite{borlandbouchaud, Mandelbrot}. Indeed that model is a variation of the Nobel-prize winning GARCH family \cite{engle}, very similar to a process known as FIGARCH \cite{figarch,zumbach}.   In this paper, we shall adopt the multi-timescale feedback model as a proxy for the dynamics of returns, over time.

\begin{figure}[t]
\label{fig1}
\includegraphics[width = 4.5in,angle=90]{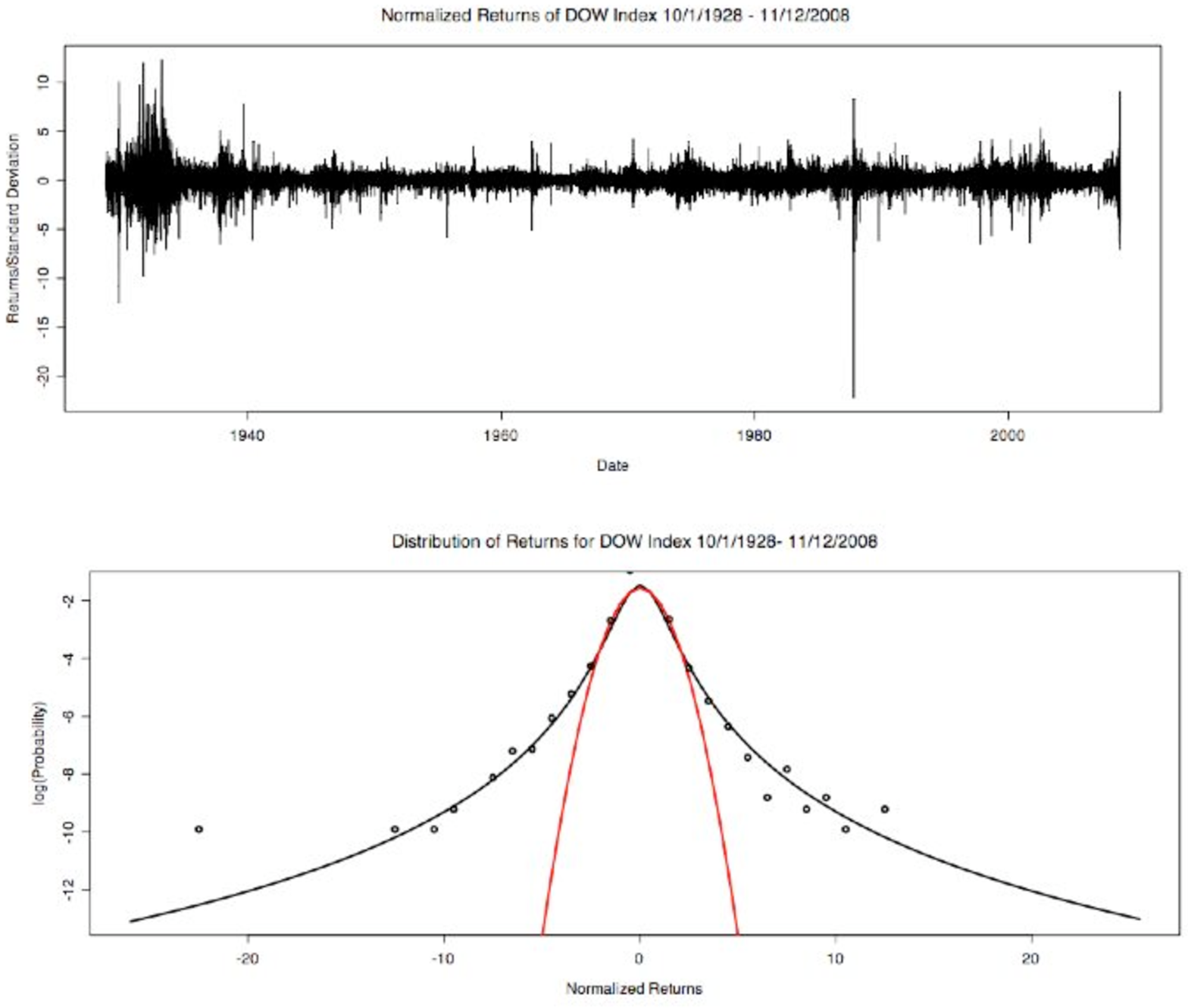}
\caption{ Returns of the Dow Jones index since the beginning of the century, together with their empirical distribution (points). The lines correspond to a Gaussian fit to the data which underestimates the tails (red), as well as a Student distribution with 5 degrees of freedom (black).}
\end{figure}

\section{Cross-sectional properties of returns}

We have just  reviewed the rather universal  time series behavior of stocks,  but now turn our attention to focus on the cross-sectional behavior of stocks, over time.  Other studies of cross-sectional market data and behavior have been presented in the literature \cite{variety, kaizoji, sornette, marsili, marsili1}, but our goal here is to see if there are any particular statistical signatures in periods of market panic. How we define market panic will become clearer later on, but  we do know apriori that we expect the current period (2008-2009) to be one of panic, as well as perhaps the years leading in to  the 2002 which correspond to the bursting of the dot-com bubble,  as well as the crash of 1929.  In the current paper we restrict our analysis  to the time period 1993 - 2009. To get a grasp on the cross-sectional distribution of stock returns at  a given time point, we can look at moments such as the mean, standard deviation, skew and kurtosis. We then look at these as a function of time.  The standard deviation of returns is widely referred to as dispersion. Calculated across a universe of 1500 US stocks  and plotted out for the time period 1993 - 2009 (Figure (2)), it is striking to see that the dispersion gets relatively big during the time periods defined as panic according to the discussion above. However, the more striking discovery is to plot out the cross-sectional kurtosis alongside the dispersion (Figure (3)), or together with market returns (Figure (4)). Even by eye it is quite clear that there is a strong negative correlation between the two quantities, which is in fact about $-25\%$. In times of panic, dispersion is high yet excess kurtosis practically vanishes. In more normal times, the dispersion is lower but the cross-sectional excess kurtosis is typically  very high.

\begin{figure}[t]
\label{fig2}
\includegraphics[width=4.5in]{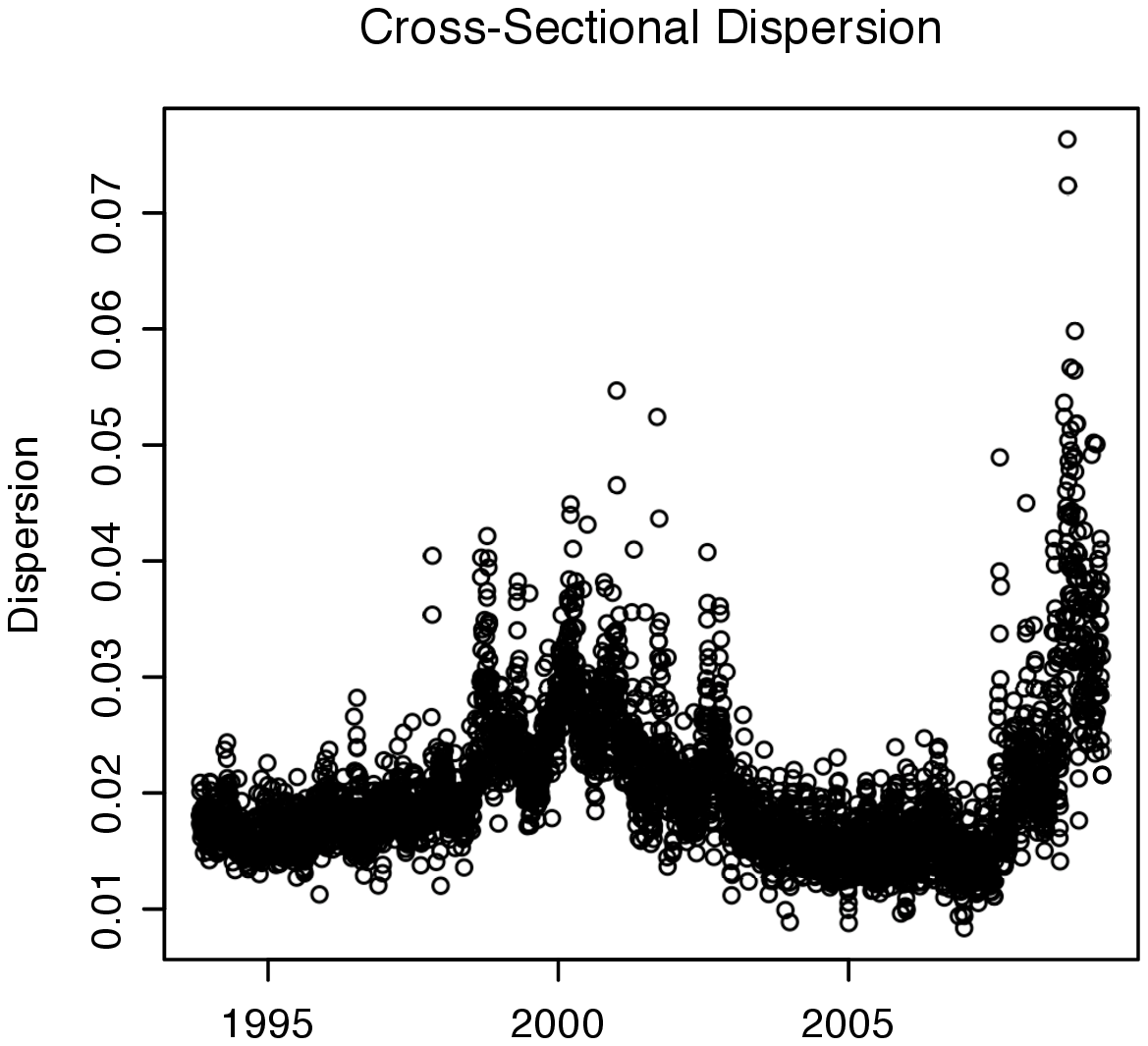}
\caption{ Cross-sectional dispersion calculated for a universe of 1500 US stocks.}
\end{figure}

\begin{figure}[t]
\label{fig3}
\includegraphics[width=4.5in]{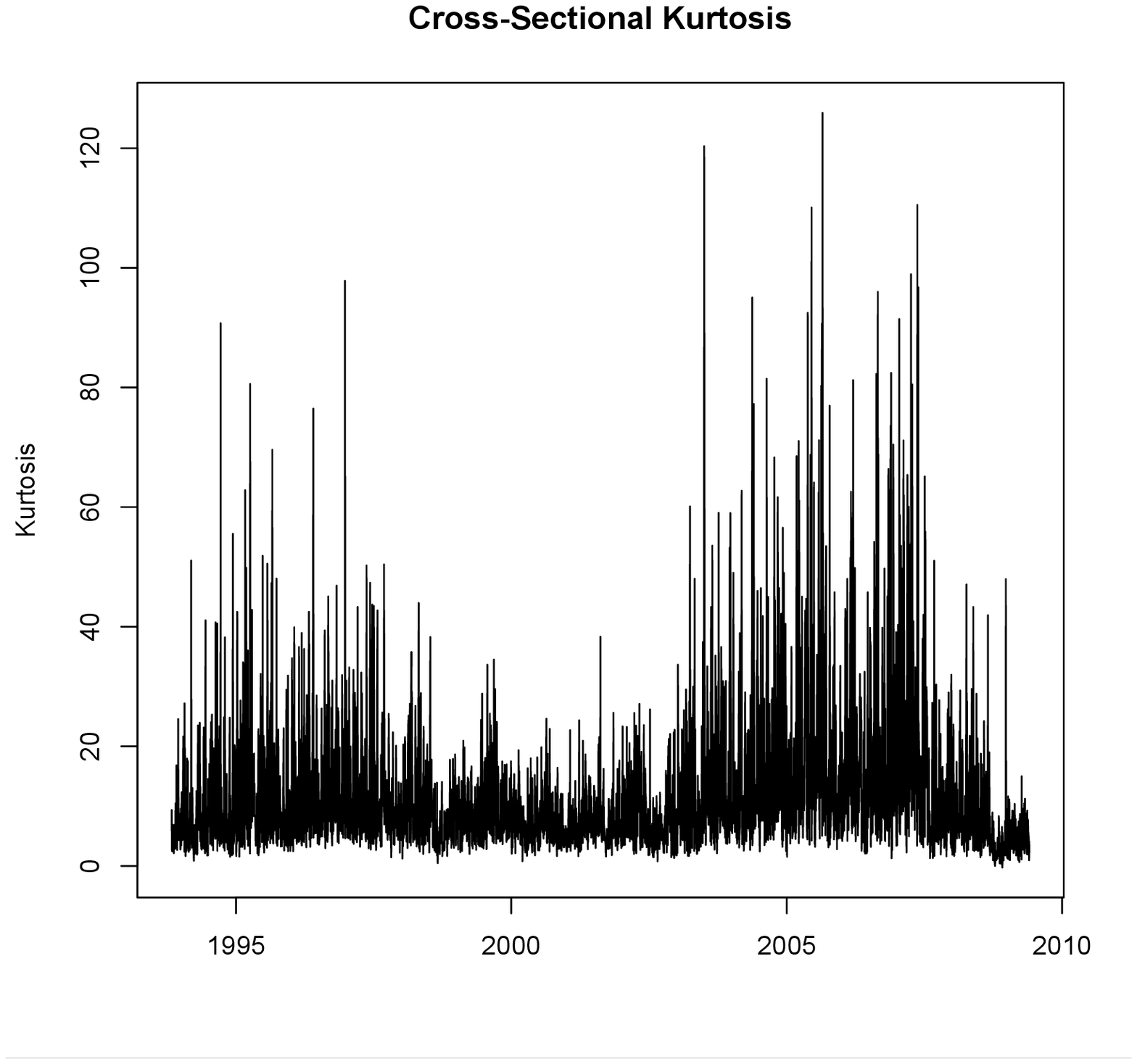}
\caption{Cross-sectional kurtosis calculated for a universe of 1500  US stocks.}
\end{figure}

This finding is at first sight counter-intuitive. Somehow we associate market panic with rare events and a wild distribution with high kurtosis. Instead we find a more Gaussian cross-sectional distribution.  The skew and the mean are also correlated with the kurtosis, but in what follows we pay less attention to these moments. The paper will mainly be concerned with understanding the dynamics of dispersion and cross-sectional kurtosis, along with the properties of correlation to which we shall turn our attention  next.

\begin{figure}[t]
\label{fig4}
\includegraphics[width=4.5in,angle=90]{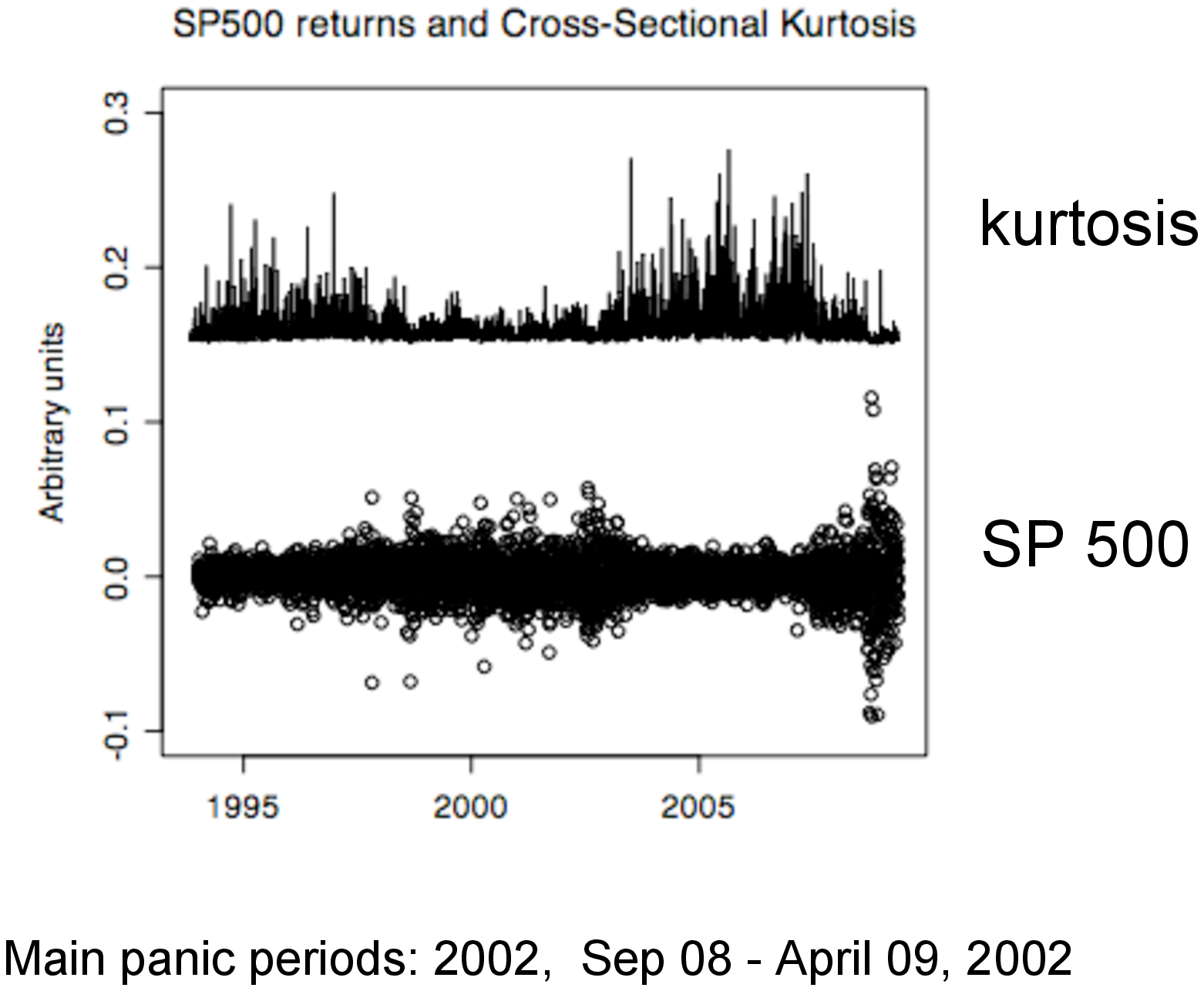}
\caption{ In the years that we study, there are two market panic times. One is 2002 (the burst of the dot-com bubble), and late 2008-April 2009. Market volatility spikes, cross-sectional dispersion rises, and cross-sectional kurtosis drops in those times.}
\end{figure}

To get a  picture of what is going on with correlation across stocks, we perform rolling principal components analysis (PCA) across the universe of stocks, with a 100-day look back window in time. Each day we plot out the  spectrum of eigenvalues, in particular we show the percentage of variance captured by the first principal component over time. 
Our hypothesis is that the larger the percentage of variance captured by this factor, the more we believe that  a "market" model exists, or in other words the larger is the co-movement of stocks. We repeated this analysis also for the volatility (defined  simply as the absolute value of the return), and changes in the volatility, plotted out here for a universe that was restricted to the SP 100  stocks (Figure(5)). In all cases the results tell the same story: there is a dramatic increase in co-movement of all of these quantities during periods of market panic, in particular during the current crisis but also in 2002.

\begin{figure}[t]
\label{fig5}
\includegraphics[width=4.5in,angle=90]{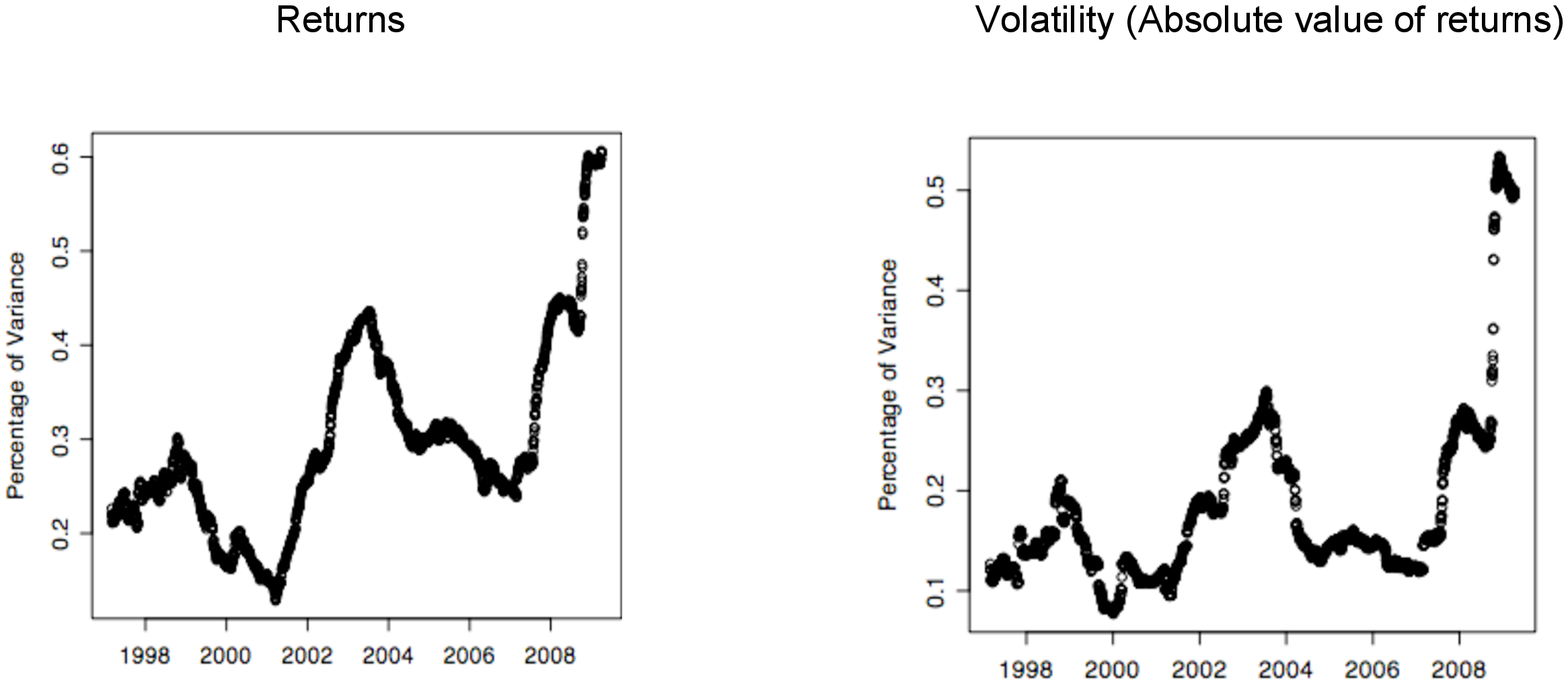}
\caption{ Percentage of variance captured by the first eigen-value for returns and volatility, calculated for the SP100 universe with a  100 day rolling window.}
\end{figure}

\section{A joint stochastic model }

Our quest now is to come up with a model that can explain all of these findings, namely to preserve the fat-tailed time series properties of stocks, yet to attain the remarkable reduction in kurtosis and increase in correlations, cross-sectionally, in particular. Let us first try and understand the changes in cross-sectional distribution. The statement that "dispersion is high yet kurtosis is low" implies that the data are more Gaussian in time of panic (see Figure (6)) and  could be explained partially by the fact that  the volatilities of the individual stocks are higher yet more alike in times of panic, a statement that is borne out by the PCA analysis of volatilities. The quantity to study is the volatility of the volatility, normalized by the mean volatility. If this is high, then the distribution of volatilities will be heavier tailed. If this is low, then the distribution of volatilities will be narrower. A plot of this quantity from market data shows that it dips in times of panic (Figure(7)).  Clearly then, if in normal times volatilities  are more diverse, then the cross-sectional distribution of returns will be  the superposition of random variables each drawn form its own distribution with own volatilities, all quite dissimilar. The resulting cross-sectional distribution can be seen as if the random variables were drawn with stochastic volatility and you would expect a fat-tailed result. If, on the other hand, all volatilities are more alike, the resulting distribution should look as if each return was drawn from the same Gaussian distribution with very similar volatilities. Note that we assume a Gaussian distribution because the driving noise of each stock time series is a Gaussian random variable, uncorrelated in normal times.
Simulations of random variables drawn from distributions where the cross-sectional volatility of the volatility normalized by the mean volatility varies from 0.8 to 0.1 are shown (Figure(8)), and indeed the distribution of cross-sectional returns becomes more Gaussian as the volatility distribution narrows.
\begin{figure}[t]
\label{fig6}
\includegraphics[width=4.5in,angle=90]{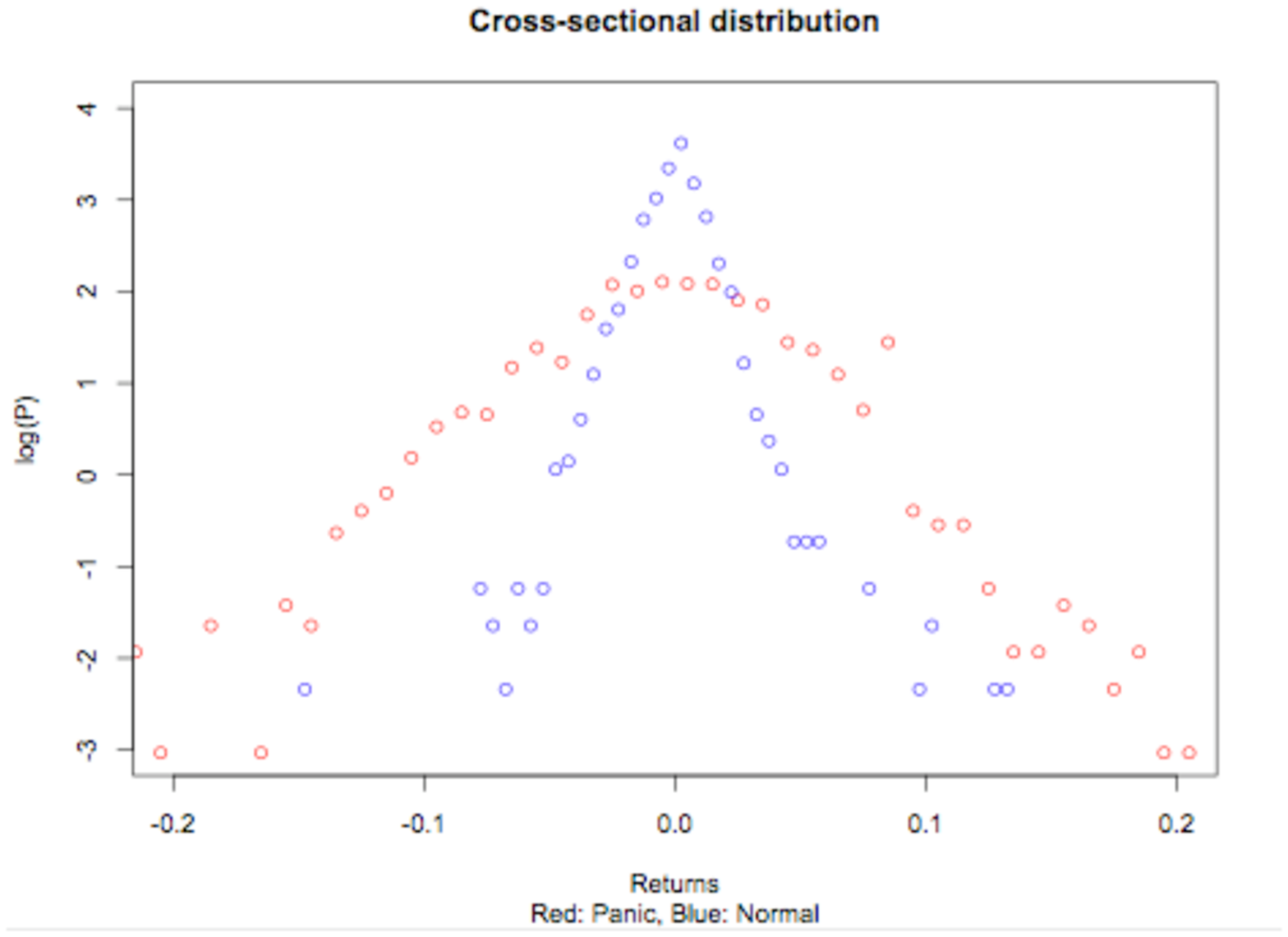} 
\caption{Empirical distributions of returns for a typical panic day are more Gaussian than on  a typical normal day, but with a larger standard deviation.}
\end{figure}
\begin{figure}[t]
\label{fig7}
\includegraphics[width=4.5in,angle=90]{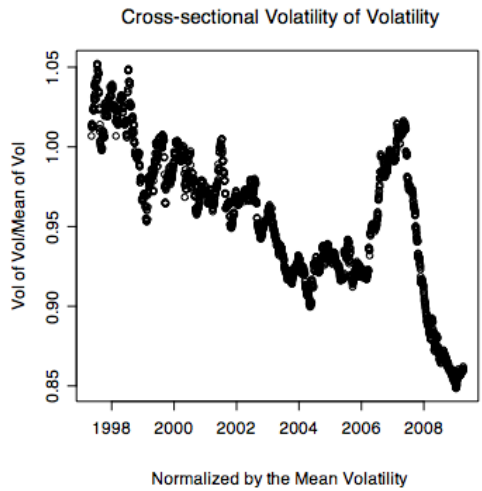}
\caption{ The moving average of the ratio of the  standard deviation of 
               dispersion to mean dispersion shows that in times of panic the distribution of cross-sectional  volatility narrows.}
\end{figure}

\begin{figure}[t]
\label{fig8}
\includegraphics[width=4.5in,angle=90]{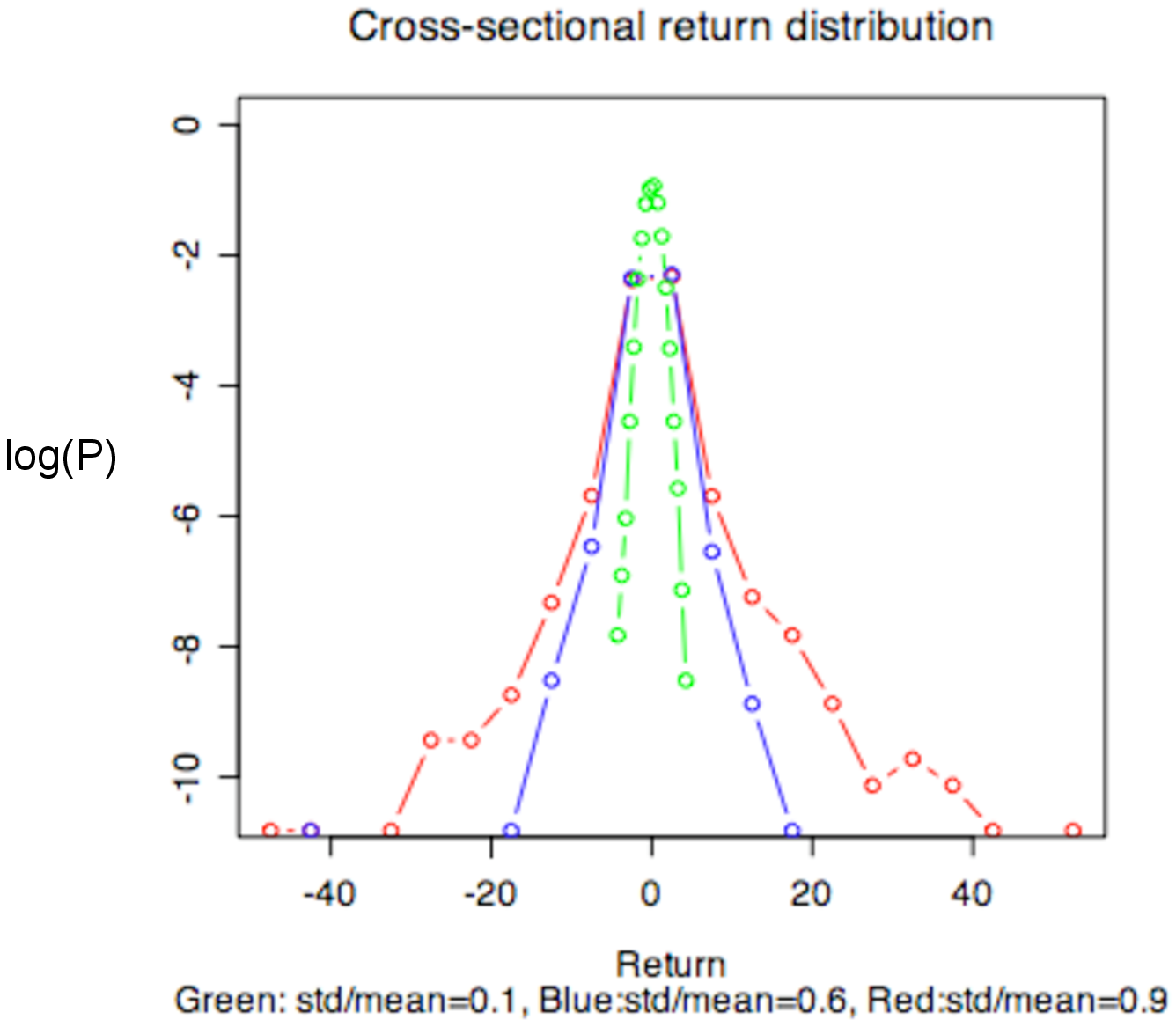}
\caption{Kurtosis gets smaller as cross-sectional volatilities become more similar.}
\end{figure}

This might be one effect contribution to our findings, but  we believe that the behavior of cross-sectional correlations holds the key to understanding the statistical signatures  that we found.  To this end, we define the following quantity \begin{equation} s = \frac{ s_{up} - s_{down}}{ s_{up} + s_{down} }\end{equation} where $s_{up}$ is the number of stocks that have positive returns over a given interval, and $s_{down} $is  the number of stocks that have negative  moves on that same interval (for example a day). If $s=0$ then roughly the same number of stocks moved up as down, and the assumption is that the stocks had little co-movement and so were uncorrelated. If all stocks move together either up or down, though, the s will be +1 or -1 and the stocks will have high correlation.  So, the following picture emerges: If $s= 0$ there is no correlation, and we are in a disordered state. However if $s \ne 0$ then there is correlation and we are in an ordered state.  We will now make a leap and borrow some terminology from physics. We shall call $s$ the  order parameter. It is a macroscopic parameter that tells us whether there is  order and correlation in the system, or not. In physics, in particular in the field of non-equilibrium thermodynamics and synergetics \cite{haken},  the concept of the order parameter is often used to describe systems that exhibit spontaneous self-organization. Examples range from chemical kinetics to laser dynamics, from fluid dynamics to biological systems; from collective behavior in both the animal and human world to cloud formation.  To illustrate the concept, let us look at an example which should be familiar and intuitive to most, namely magnetism.

\subsection{Ferromagnetic dynamics}

In a ferromagnetic system,  the total magnetic moment depends on the orientation of the individual magnetic spins comprising the system.  It is proportional to the quantity
\begin{equation} m= \frac{ m_{up} - m_{down}}{ m_{up} + m_{down} }\end{equation} where $m_{up}$ and $m_{down}$ denote the number of spins lined up and down respectively. The distribution of possible outcomes  of this macroscopic quantity is given by \begin{equation} P(m) = N \exp ( F(m,T)) \end{equation} where $T$ is the temperature, $N$ is a normalization factor and $F$ is the free energy of the system. The temperature is an important parameter in this system because as we shall see, depending on the value of $T$, the magnetic system will either be in an ordered or discorderd state. The next step is to perform a Taylor expansion of $F$ where symmetry arguments are used to eliminate the first and third order  terms, yielding
\begin{equation}  F = a m^2  + b m^4. \end{equation}  where $a$  and $b$ are parameters.  Every probability distribution can be associated with a  Fokker-Planck equation describing its temporal evolution. The Fokker-Planck equation in turn is associated with  a Langevin equation, describing the dynamics of the underlying variable itself. In the case of the magnet, the corresponding Langevin equation takes the form  \begin{eqnarray} \frac{dm}{dt}&  =  & - \frac{\partial{F}}{\partial{m}} + F_t \\&= & -\frac{a}{2} m - \frac{b}{4} m^3 + F_t \end{eqnarray} where $F_t$ is thermal noise. Now, one writes \begin{equation} a = \alpha(T-T_c) \end{equation} where $T_c$ is the so-called critical temperature.  One can envision these dynamics as motion in a potential well  $V$ given by $ V(m) = -F(m)$, as is illustrated in the Figure (9). It is easy to see that if $T>T_c$, the only minimum is the trivial one at $ m = 0$. However,  for $T < T_c$ there are two real roots appearing, yielding non-zero values of $m$.  Clearly, $m$ can be positive or negative, depending on which minima is reached by the system. This is referred to as symmetry breaking. Due to the noise, the dynamics can also drive  $m$ from one minimum to the other.  Because the value of $T$ determines whether the system is in the disordered state ($m= 0$) or the ordered state ($ m \ne 0)$, it is called the control parameter. The probability distribution of the system in the disordered state will be a unimodal one, while the probability distribution of $m$ in the ordered state will be bimodal.  As $T$ passes from above to below $T_c$, or vice-versa, there is clearly a phase transition:  the state of the system is drastically altered. In this type of symmetric system, the phase transition is referred to as a second order one.

\begin{figure}[t]
\label{fig9}
\includegraphics[width=4.5in,angle=90]{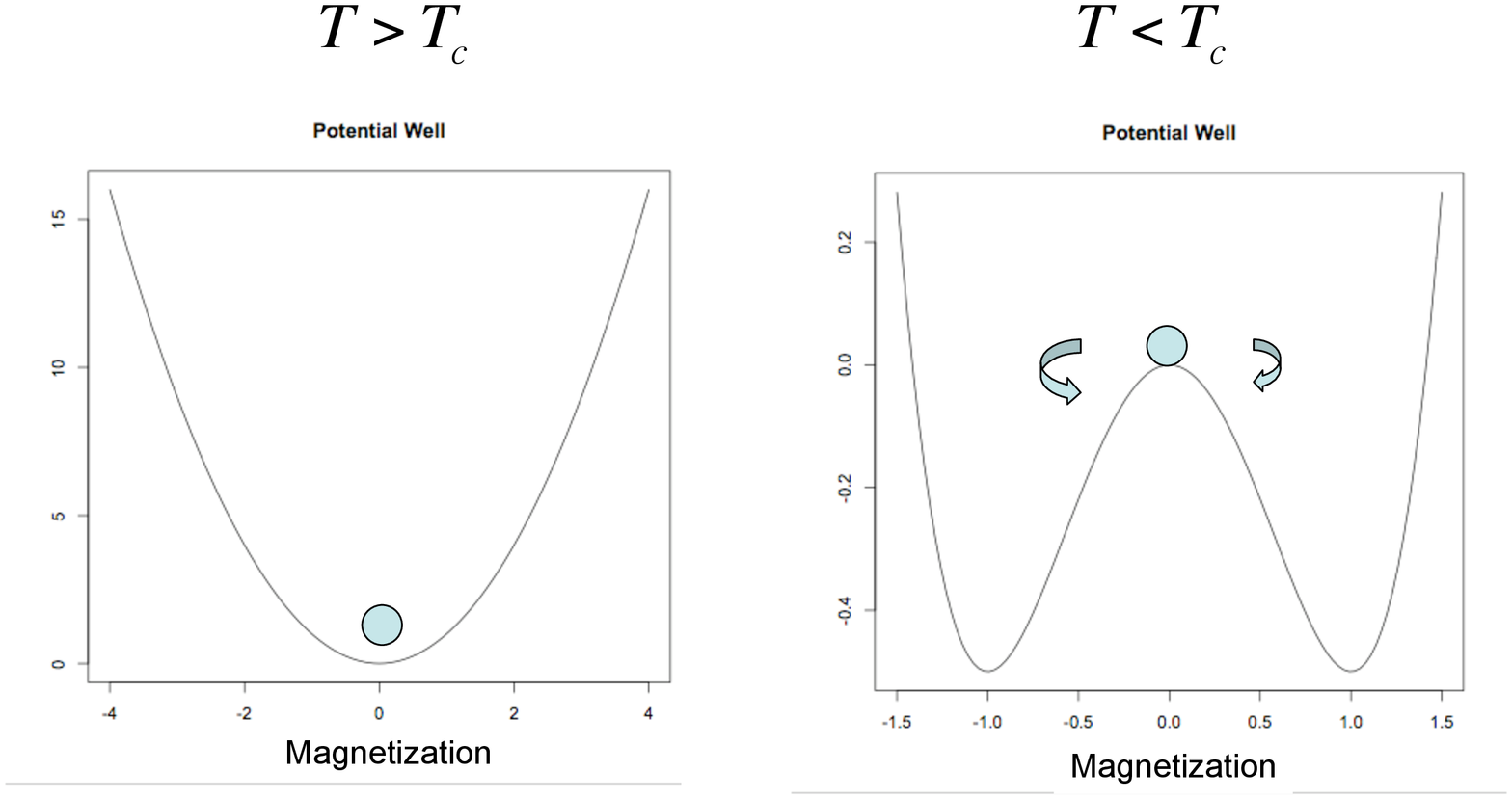}
\caption{ In a ferromagnetic system, a  phase transition is induced when the control parameter $T$ drops below a critical temperature $T_c$.}
\end{figure}

\subsection{Self-organization of correlation  and a  cross-sectional model of returns across time }

Now what has this got to do with our system? Our variable $s$ strikes a similarity to $m$. We have observed rather drastic changes in the cross-sectional distribution of stocks in the times of panic versus more normal market conditions.  Let us look at   histograms of  $s$ in both periods, as shown in the Figures (10) and (11). It is quite clear that in normal times, $s$ is unimodal, and in panic times we obtain a bimodal distribution consistent with the frame-work of a phase transition leading to self-organization in panic times.
\begin{figure}[t]
\label{fig10}
\includegraphics[width=4.5in,angle=90]{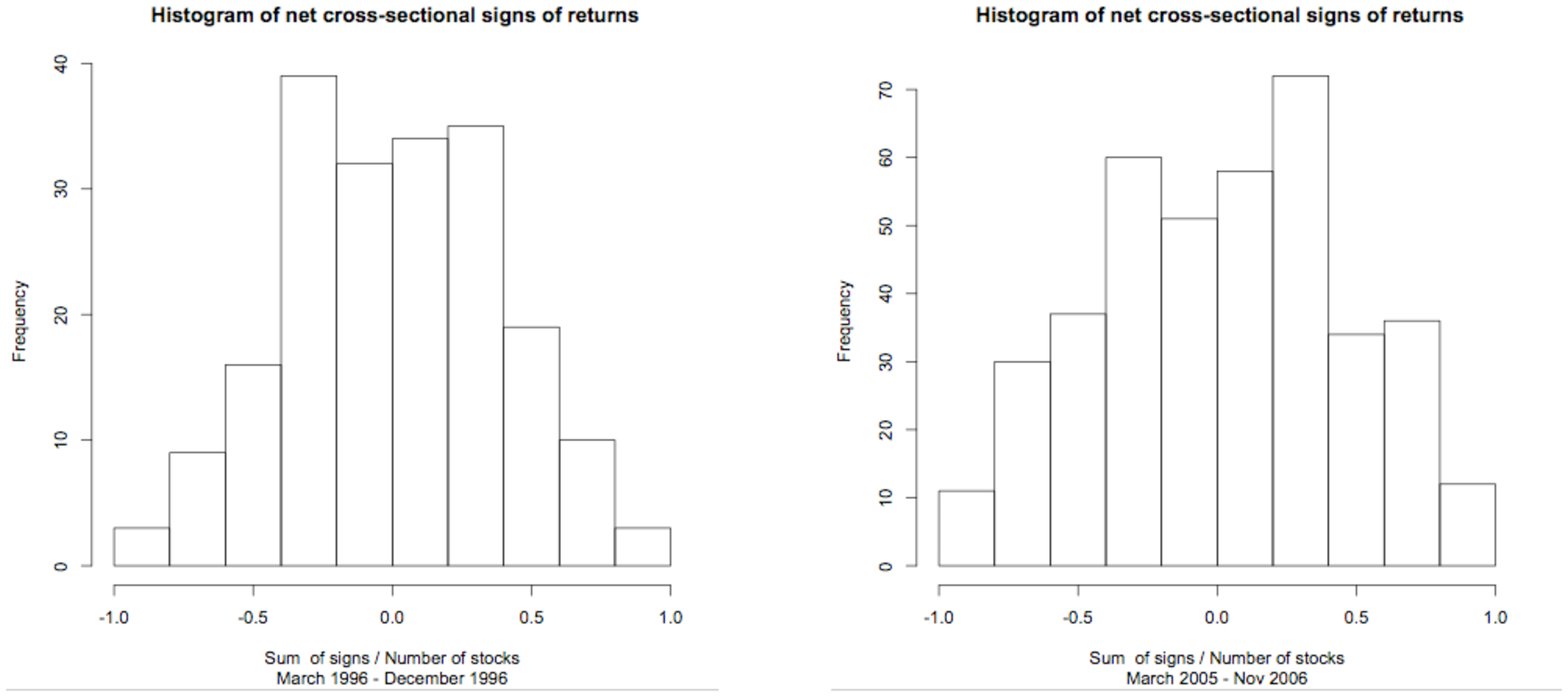}
\caption{ In the US financial market, $s$ (the normalized sum of the signs of returns) has a unimodal distribution in normal times.}
 \end{figure}

\begin{figure}[t]
\label{fig11}
\includegraphics[width=4.5in,angle=90]{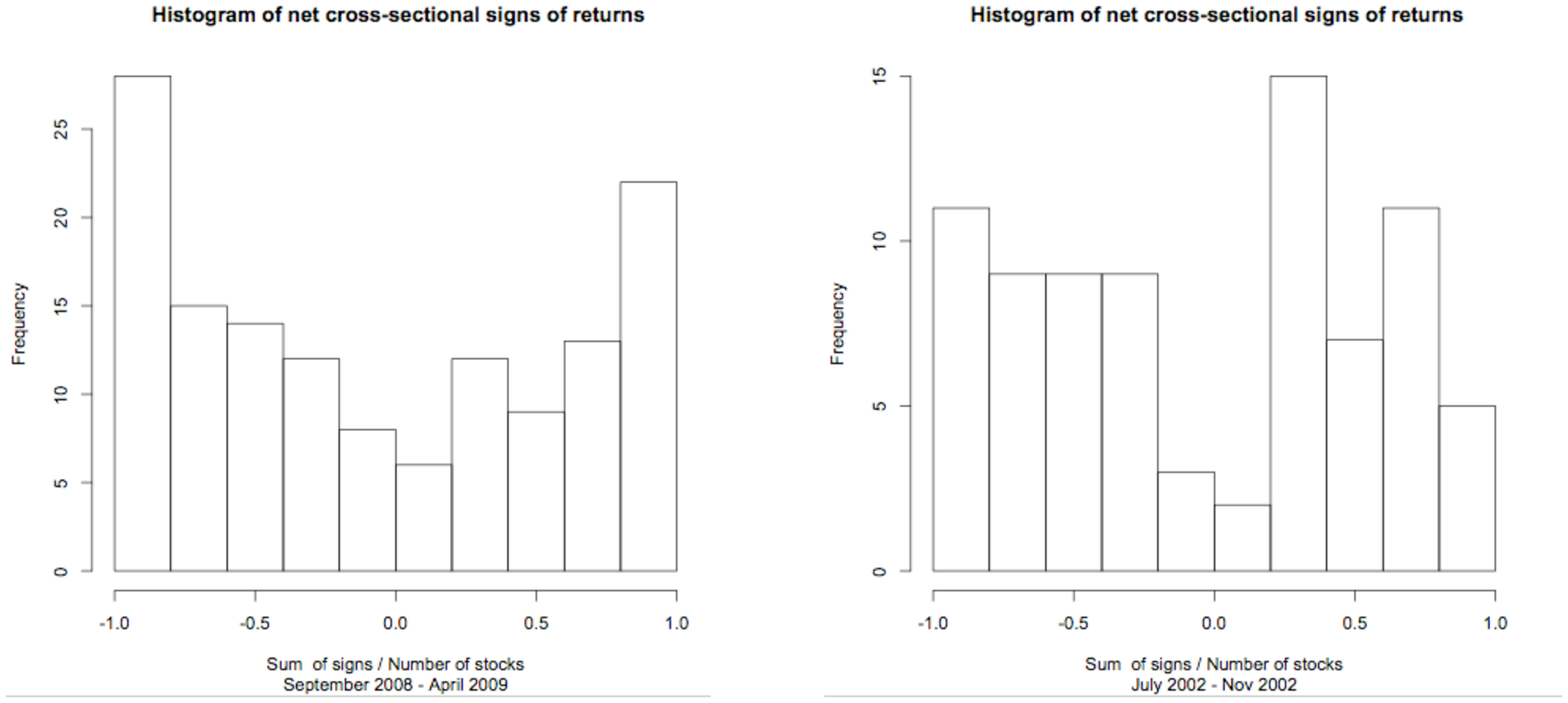}
\caption{ In the US financial market, $s$ has a bimodal distribution in panic times.  In our model, $s$ plays the role of the order parameter, akin to the magnetization in a ferromagnetic system. It is closely related to the correlation across stocks.}
\end{figure}

 We postulate that the dynamics of $s$ be given by
\begin{equation} \frac{ds}{dt} = -\frac{a}{2} s  - \frac{b}{4} s^3 + F_t .\end{equation}
We propose that \begin{equation} a =  \sigma_c - \sigma_0 \end{equation} where $F_t$ is a Gaussian noise term and $\sigma_0$  corresponds to the  baseline volatility level of stocks which will be discussed in more detail below. This volatility is assumed constant across all instruments, and essentially measures the general  uncertainty in the environment, so in this sense acts much as the temperature in the magnetic system.  Note that it is the feedback effects  in the system which induce stock-specific variations in volatility over time,  and can largely explain most of the excess volatility observed in stock time-series, whereas the parameter $\sigma_0$ is not driving the stock-specific dynamics, but simply describes a "global" level of risk.  The quantity $\sigma_c$ would correspond to a critical level of uncertainty, below  which the market is in a normal phase, and above  which we have the onset of panic. Much as in the case of ferromagnetism, where the control parameter $T$ can be tuned  externally above or below the critical temperature, in our model the uncertainty level $\sigma_0$ captures the external  environment. In a sense it represents the general perception of risk in the public mind.
Our hypothesis is then that  financial markets appear to  exhibit a phase transition from the disordered to ordered state, after crossing a critical level of risk perception.  Putting the dynamics together, we have the multi timescale feedback process for each stock \cite{borlandbouchaud}, \begin{equation}
 d y^k_{i+1} = \sigma^k_i  d \omega^k_i
\end{equation} with

\begin{equation}
\sigma^k_i = \sigma_0 \sqrt{ 1 + g \sum_{j = 1}^{\infty}\frac{1}{(i-j)^{\gamma}} (y^k_i  - y^k_j)^2}
\end{equation}
where $k$ runs over all N stocks and $i$ corresponds to time. The parameter  $g$  is a coupling constant that controls the strength of the feedback, $\sigma_0$ is the baseline volatility discussed above, and $\gamma$ is a factor that determines the decay rate of memory in the system. In this formulation we assume a unit time-step. In  \cite{borlandbouchaud} a more general  formulation which includes the continuous time limit is shown.
The random variables $\omega^k_i$ are drawn from a Gaussian distribution, uncorrelated in time such that  $ < \omega^k_i \omega^k_j > = \delta_{ij} $ , yet amongst themselves at a given time point $i$ across stocks $k$, they are correlated  with a correlation $|s|$.  The macroscopic order parameter $s$  is therefore just  a signature of the cross-stock  correlations, whose dynamic behavior  manifests itself in the order parameter equation \begin{equation} \label{eqs} \frac{ds}{dt} = - \frac{a}{2} s - \frac{b}{4} s^3 + F_t .\end{equation}   For numerical simulations, we cannot really run the dynamics of the normalized sum of signs $s$ directly since it is in fact the dynamics of the underlying correlation $|s|$ of the Gaussian random variables that we need to simulate. Another caveat is that the  coefficients must always be such  that $|s| \le 1$.  We solve these problems by obtaining $|s|$ from an equation of the same form as Eq(\ref{eqs}) such that $|s| =   |\tanh(\hat{s})|$.

What do we expect to see from this model? Across time,  this process is stable  if 
$ \sum_{j = 0}^{\infty}\frac{g}{i-j}^{\gamma} <1  $ or approximately that $\frac{g}{1-\gamma} < 1$ and $\gamma > 1$ \cite{borlandbouchaud}.
It captures the known stylized facts if $ \gamma = 1.15$ and  $g = 0.12$, as is discussed in detail in  \cite{borlandbouchaud}.  For illustration we show in Figure (12) a  simulation of a truncated version of that model with only  30 days of  memory just to point out that the   process describes real data very well, with obvious periods of lower and higher volatility clustering together.  The fat-tailed distribution of the simulated returns is also shown.

\begin{figure}[t]
\label{fig12}
\includegraphics[width=4.5in,angle=90]{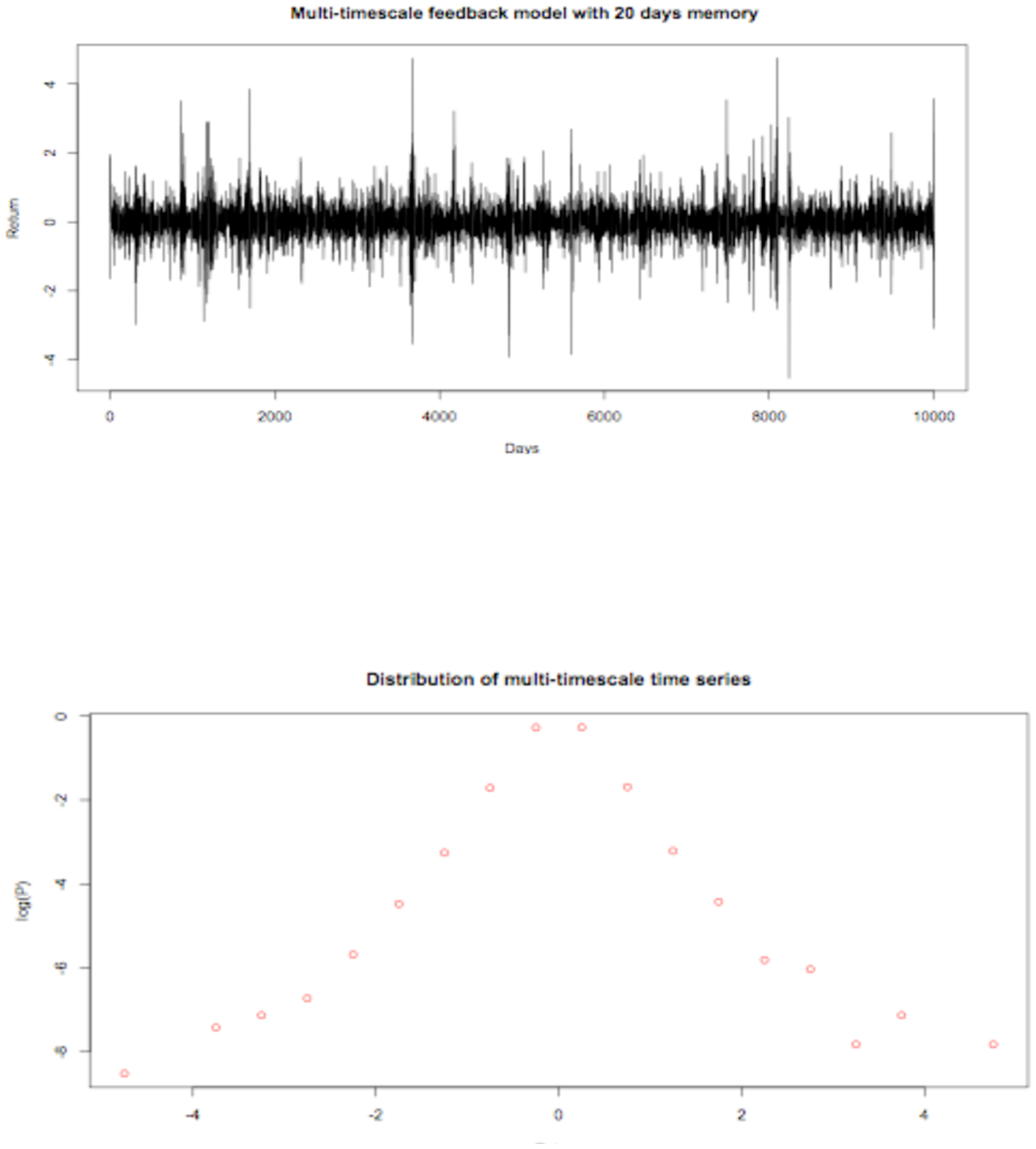}
\caption{ A simulation of returns generated by the multi-timescale model with 30 days of memory, together with the corresponding distribution of returns.}
\end{figure}

Across stocks, if $ \sigma_0 < \sigma_c$, correlations fluctuate around $s = 0$ and we expect to see a unimodal distribution of $s$. The cross-sectional kurtosis should be rather high since there is no mechanism to cause either stocks or stock volatilities to have any co-movement at all, so at each time point it is as if the cross-sectional  returns are drawn from a Gaussian process with stochastic volatility, yielding a fat-tailed distribution as the superposition. Then as the market crashes with $ \sigma_0 >  \sigma_c$, the system enters a phase transition. The order parameter $s$ becomes $s \ne 0$ and the system enters the ordered phase with high co-movement. Because the random variables $\omega^k_i$ are now correlated across stocks, cross-sectional returns will be more similar and the distribution will have lower kurtosis. Additionally, due to the fact that the phase transition is triggered by an external shock in volatility, all stocks  will tend to have higher volatilities and higher cross-sectional dispersion. Note that if the phase transition were instead triggered by an extreme negative return (not a volatility shock),  due to the  feedback mechanism, a transition of $s$ from the disordered to ordered phase would reduce the kurtosis but also  reduce  the dispersion because although stock volatilities would increase, they would all move very much together since the terms in the  feedback mechanism  would all be rather similar.  In fact we expect the volatilities to be more coherent as dictated by the feedback, but at a higher volatility level, which is accounted for by the increase in $\sigma_0$.

We point out  that our findings are consistent with the idea of market panic as a phase transition, which was proposed by Bouchaud and Cont  \cite{bouchaudcont} in a different setting. The common theme between  the  two approaches includes the importance of feedback effects. We are also aware  that work has been done on Ising models of markets \cite{sornette,Sornette}, 
typically modeling the interaction of  participants (or agents) similar to those in a magnetic system. Here again our approach is different,  as we describe the dynamics on the macroscopic level of an order parameter $s$. However,  it would be really interesting  to make a connection between microscopic (perhaps Ising-like)  and macroscopic dynamics. 

\section{Numerical Simulations}

Simulations of this model for the joint stochastic process of stocks were performed. To keep things simple and quicker to simulate, we  implemented a simpler form of the multi-timescale model which included memory over only 30 past time steps. We have not calibrated the parameters to represent realistic values of returns - we simply want a toy model that reproduces some main features of volatility clustering and  a fat-tailed distribution of the resulting time-series.  The baseline volatility $\sigma_0$ was chosen to equal $\sigma_b = 0.20 $  which is a realistic assumption.  The driving noise for each stock's time series was obtained by running the  dynamics for the correlation \begin{equation} \frac{d\hat{s}}{dt} = -(\sigma_c - \sigma_0)\hat{s}- b\hat{s}^3 + \hat{F}_t \end{equation} where $b = 0.01$ and $\sigma_c$ was chosen to be  $\sigma_c = 0.4$ or twice  the usual base volatility.  The noise term $\hat{F}_t$ was drawn from a zero mean Gaussian distribution with a standard deviation of $0.1$. Based on the value of $\hat{s}$, a random Gaussian correlated noise for each stock was calculated using  the Cholesky decomposition with  a correlation equal to $|\hat{s}|$ . At $t = 250$ in our simulation, a large volatility shock was applied to the system such that \begin{equation} \sigma_0 = \sigma_b + \sigma_{shock} \end{equation} with $\sigma_{shock} = 0.6 $, which is a rather realistic choice if one remembers levels of the VIX in late 2008. This induced the phase transition  from the disordered state where correlation among stocks are relatively low, centered around zero,  to a highly ordered state where the correlations are different from zero.

Our results show that the main features of financial markets are  captured within this framework. The correlation  $|s|$  goes from  $0$  (the disordered state) to  $|s| \approx 0.8$ ( the ordered state) at the time of the volatility shock. When the shock subsides, it returns to  the disordered state again. 
A plot of the mean superposition of 200 realizations, which represents the market is shown in Figure (13). As expected,  the market volatility rises  when  $s$ is in the ordered state, which corresponds to the panic phase. In addition, Figure (14) shows that  the cross-sectional dispersion rises during the market panic,  while the cross-sectional kurtosis drops close to zero (Figure (15)). The correlation between the two quantities is in this example $-.17$, consistent with empirical observations that also showed a strong negative correlation.  We also plotted out histograms corresponding to the distribution of the order parameter $s$ in the normal market phase as well as in the panic phase (Figure (16)). These distributions are in excellent agreement with the  empirical observations of the real market data, namely unimodal in the normal phase, and clearly bimodal during the panic time.

\begin{figure}[t]
\label{fig13}
\includegraphics[width=4.5in]{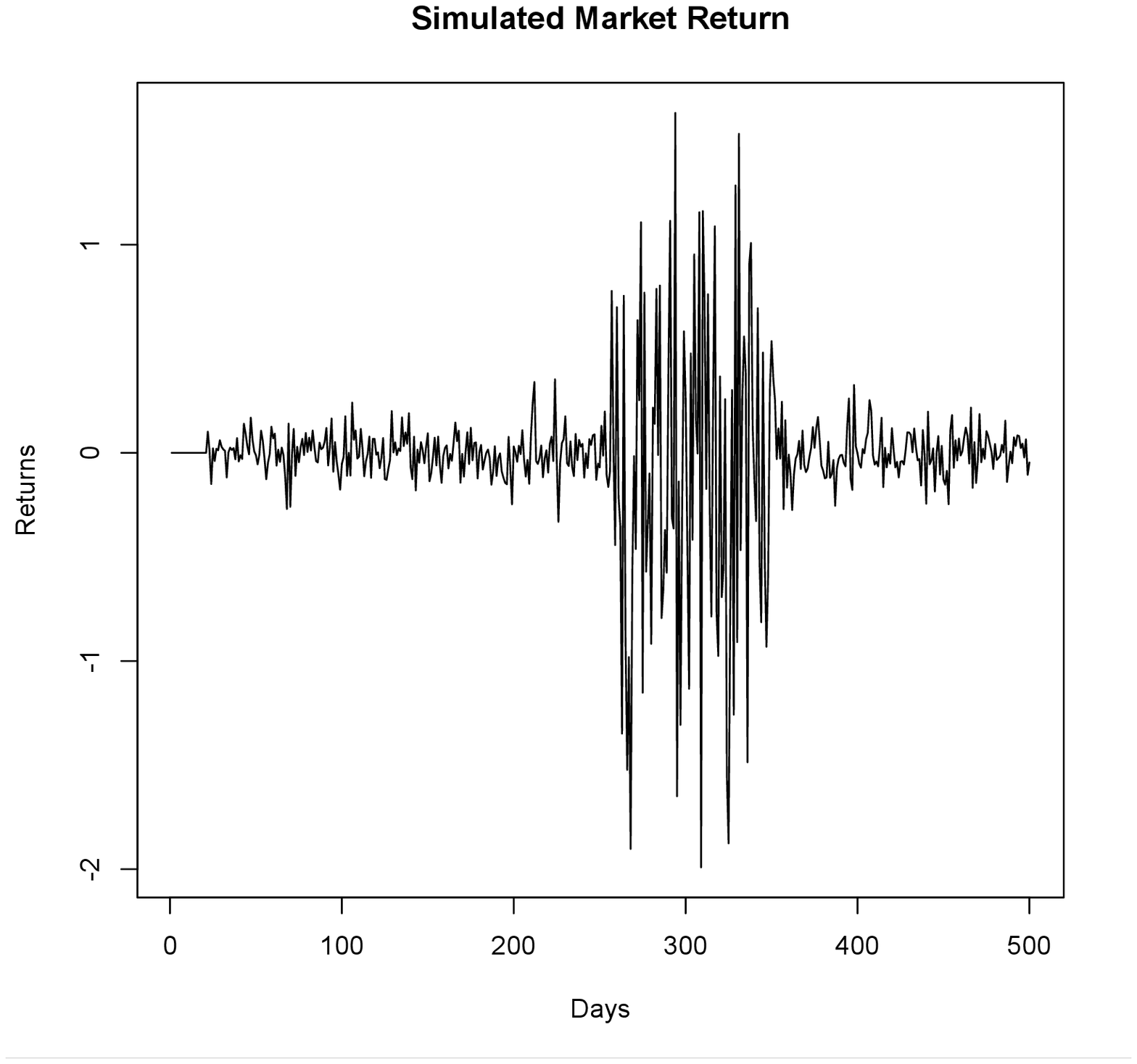}
\caption{Market volatility rises in the  panic time, induced at time $t=250$ when $\sigma_0 > \sigma_c$.}
\end{figure}

\begin{figure}[t]
\label{fig14}
\includegraphics[width=4.5in]{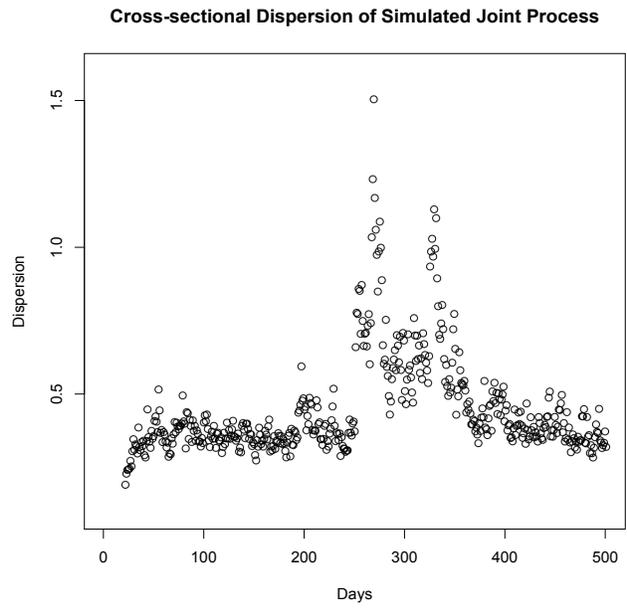}
\caption{ At the onset of panic ($t= 250$), cross-sectional dispersion increases markedly as a signature of the phase transition.}
\end{figure}

\begin{figure}[t]
\label{fig15}
\includegraphics[width=4in]{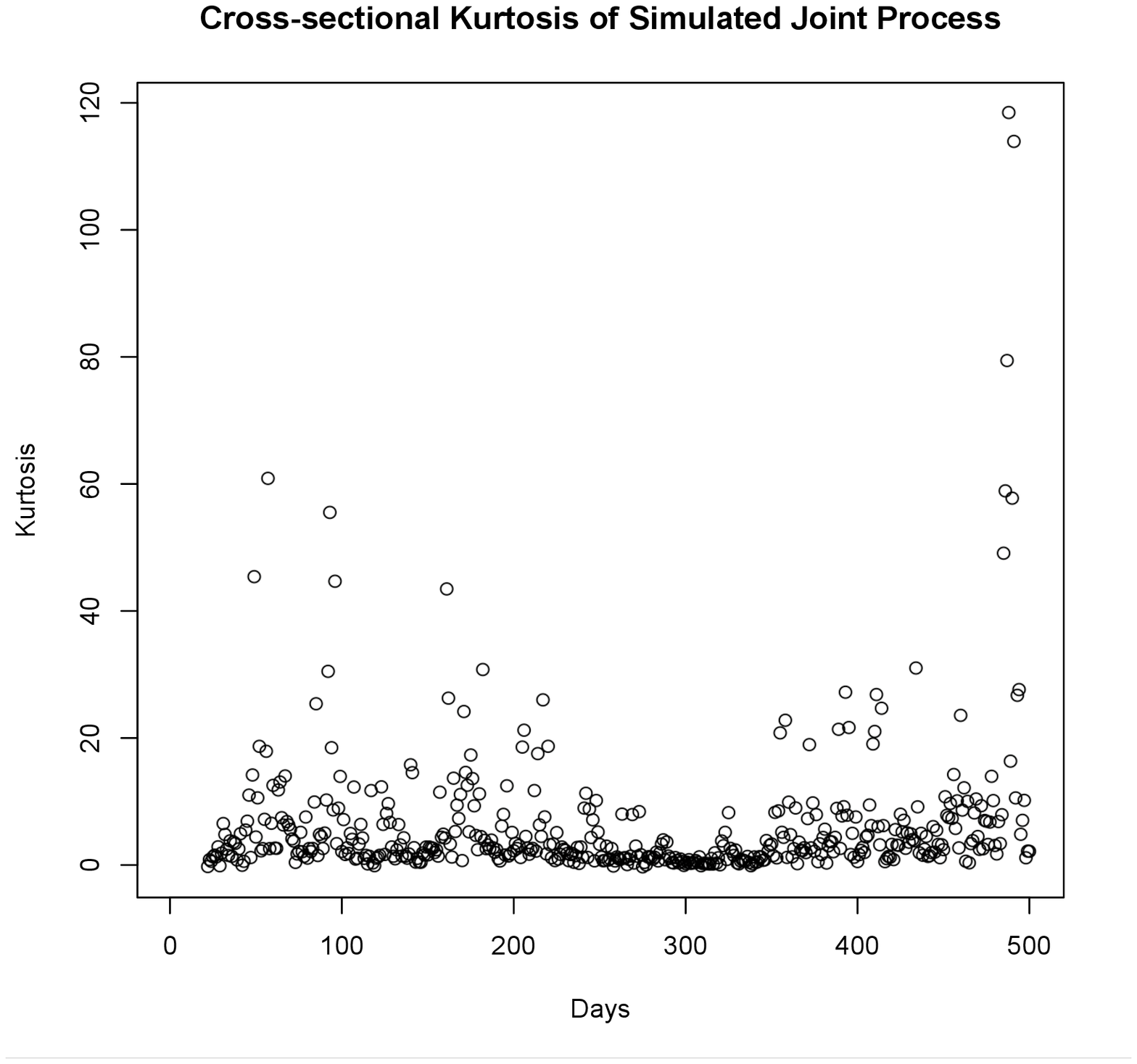}
\caption{Cross-sectional excess  kurtosis drops close to zero in times of panic ($t=250$ to $t=350$).}
\end{figure}

When the volatility shock that was applied to the system at some point dies away (in our example this happens around t = 350), the market undergoes another phase transition to the disordered state.  An obvious question pertains to the duration of the volatility shock, and how does it subside?  In simulations we looked at different scenarios.  We applied a constant shock that lasted a certain number of days and then was turned off.  This would correspond to a market environment where there is persisting uncertainty in investor sentiment.  Even after the shock was turned off, its effects would linger in the system due to the feedback and memory mechanism. Hence, shorter duration but large amplitude shocks could also  have longer lasting effects. Overall, our results were qualitatively robust to the actual mechanism of  the shock decay.  In future work we hope to understand more about the volatility shock process by calibrating our model to empirical data (although we realize that this may be difficult due to limited observations as there are only a few periods of panic over history).

\begin{figure}[t]
\label{fig16}
\includegraphics[width=4.5in,angle=90]{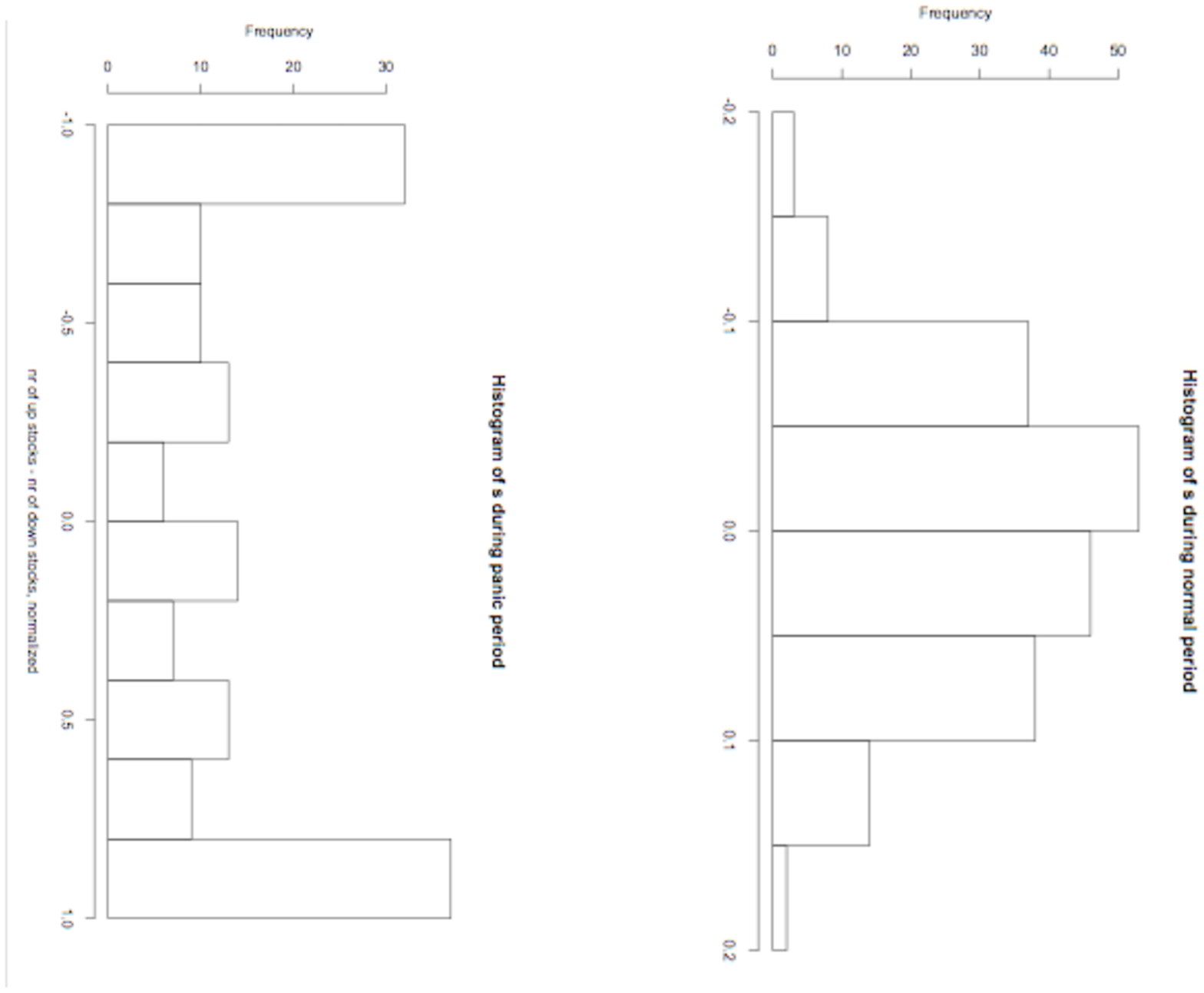}
\caption{The distribution of $s$ is unimodal in normal times, and bimodal in panic times, just as observed in the real market data. This is the signature of self-organization: the system goes from the disordered state where the most probable value of  the order parameter $s$ is $s=0$, to the ordered state  where the most probable value of $s$ is $s \ne 0$. There is symmetry breaking in that $s$ can be $\pm |s|$. }
\end{figure}

In this model, the phase transition is triggered by an exogenous volatility shock, which reproduced the observed statistical signatures of real data, in particular increasing market volatility and cross-sectional dispersion while reducing cross-sectional kurtosis.   We would like to point out that we also explored inducing the phase transition with a large negative return $r$, such that the control parameter would be of the form $(r-r_c)$ where $r_c$ would be a critically low market return, below which panic would be induced. Interestingly, that model captured all but one   of the observed statistical signatures of market panic: the market volatility rose and cross-sectional kurtosis dropped, but  dispersion also dropped and was positively correlated with the kurtosis.  The failure to capture the dispersion effect, which on the other hand is captured nicely with the volatility as the control parameter, lends weight to the current volatility-induced model in that it is rather a unique mechanism which triggers all the observed market dynamics. 
Philosophically there is a big difference between the two possible mechanisms as well. In the case of a return-driven phase transition, the system must self-induce a large enough negative return which is an endogenous effect. In the case of a volatility driven control parameter,  it is the exogenous  risk perception that will push the system into panic. This is in line with external events such as political and macro-economic factors playing a role.

\section{Summary}

In this paper we have looked at some of the properties of the cross-sectional distribution of returns over different time periods. In particular we find a significant anti-correlation between cross-sectional dispersion and cross-sectional kurtosis such that in normal times, dispersion is low but kurtosis is high, whereas in panic times dispersion is high and kurtosis is low. At first sight this appears counter intuitive as one associates panic times with wild returns, so-called Black Swans and rare events. Our finding shows rather  that the general shape of the distribution in panic times is more Gaussian, but with a higher standard deviation. So in a sense one could say that the Black Swans \cite{nassim} are simply everywhere in these times, making them in a distributional sense less rare. Another of our findings is that there is a marked increase in the co-movement of stock returns, volatilities and changes in volatility in times of panic.  We define a simple statistic, namely the normalized sum of signs of returns on a given day, to capture the degree of correlation in the system. This parameter $s$  can be seen as the order parameter of the system because if $s= 0$ there  is little correlation, whereas for $s \ne 0$ there is high correlation among stocks. 

We make an analogy to the theory of self-organization and non-equilibrium phase transitions widely used to describe collective phenomena in various fields of physics, chemistry, biology and social sciences.
 Based on this analogy we hypothesize that financial markets undergo spontaneous self-organization when the external volatility perception rises above some critical value. Indeed,  it is seen from historical market data that $s$  follows a unimodal distribution in normal times, shifting to a bimodal distribution in times of panic. This is consistent with a  second order phase transition between the disordered (normal) and ordered  (panic) states. Simulations of a joint stochastic process for the ensemble of stocks across time use a multi timescale feedback process in the temporal direction - which we know mimics the known stylized facts of time-series volatility very well -and employ an equation for the order parameter $s$ as a proxy for the dynamics of the cross-sectional correlations across stocks.  Numerical results of such simulations show good qualitative agreement with what is observed in market data.

There is a lot of future work that we would like to pursue along the lines of this model. 
Most importantly we would like to calibrate the model to market data, and see if we can indeed define a value of the volatility shock which triggers the phase transition to a panic state. Additionally, we'd like to study the dynamics of the transition back to a more normal market behavior. We also plan to  extend our current study of empirical signatures of market panic to a longer history of data, for example to take a look at the crash of 1929. Finally we want to see if there are any early warning signals of the onset of panic which would have obvious applications for trading strategies and risk control.

Acknowledgements: Special thanks to Christian A. Silva and Jeremy Evnine for many interesting discussions. Jermey Evnine is also thanked for his continual support.

\end{document}